# CellListMap.jl: Efficient and customizable cell list implementation for calculation of pairwise particle properties within a cutoff

Leandro Martínez[1]

Institute of Chemistry and Center for Computing Engineering & Sciences

University of Campinas. Campinas, SP, Brazil.

---

**Abstract**

N-body simulations and trajectory analysis rely on the calculation of attributes that depend on pairwise particle distances within a cutoff. Interparticle potential energies, forces, distribution functions, neighbor lists, and distance-dependent distributions, for example, must be calculated. Cell lists are widely used to avoid computing distances outside the cutoff. However, efficient cell list implementations are difficult to customize. Here, we provide a fast and parallel implementation of cell lists in Julia that allows the mapping of custom functions dependent on particle positions in 2 or 3 dimensions. Arbitrary periodic boundary conditions are supported. Automatic differentiation and unit propagation can be used. The implementation provides a framework for the development of new analysis tools and simulations with custom potentials. The performance of resulting computations is comparable to state-of-the-art implementations of neighbor list algorithms and cell lists, available in specialized software. Examples are provided for the computation of potential energies, forces, distribution of pairwise velocities, neighbor lists and other typical calculations in molecular and astrophysical simulations. The Julia package is freely available at http://m3g.github.io/CellListMap.jl. Interfacing with Python and R with minimal overhead is possible.

*Keywords:* cell lists, particle simulations, cutoff, neighbor list

---

[1] Corresponding author: *E-mail address:* lmartine@unicamp.br

**NEW SUBMISSION PROGRAM SUMMARY**

Program Title: CellListMap.jl

CPC Library link to program files: (to be added by Technical Editor)

Developer's repository link: http://github.com/m3g/CellListMap.jl

Code Ocean capsule: (to be added by Technical Editor)

Licensing provisions(please choose one): MIT

Programming language: Julia

Supplementary material:

Journal reference of previous version:*

Does the new version supersede the previous version?:*

Reasons for the new version:*

Summary of revisions:*

Nature of problem(approx. 50-250 words):

*Particle simulations and trajectory analysis, in chemistry, atomic and molecular physics, and astrophysics, require the computation of pairwise properties within a cutoff, for example potential energies, forces, distance distributions, among others. Cell lists can be used to accelerate these computations and are used in major simulation software, but these implementations cannot be easily adapted for novel custom calculations, simulations, and novel research.*

Solution method(approx. 50-250 words):

*This package provides a fast and customizable implementation of cell lists for the development of new simulation and analysis software for particle simulations. To provide a customizable implementation, the Julia language was used, allowing the user to write high-level yet performant custom functions to be mapped into pairs of particles satisfying the distance constraint desired. State-of-the art performance is obtained in serial and parallel executions.*

Additional comments including restrictions and unusual features (approx. 50-250 words):

*The implementation is type-generic, allowing systems in 2 or 3 dimensions to be studied, and the computations can be written in such a way to allow the propagation of units, uncertainties, and automatic differentiation of the computed properties.*

# 1 INTRODUCTION

Particle simulations are fundamental in the study of molecular and astrophysical phenomena, among other areas of research. The simulations' implementation is dependent on the calculation of particle interactions, and their analysis is dependent on the computation of different attributes from the resultant trajectories. The computation of distances between particles within a cutoff are required for many of these calculations, being frequently the most computationally expensive step.

The computation of all pairwise distances, with a cost of $O(n^2)$ where $n$ is the number of particles, rapidly becomes too expensive for practical application. As a result, techniques that prevent probing particle pairs that are too far apart to show important interactions or correlations must be utilized. The most often used algorithms are those based on distance trees or cell lists [1]. In simulations with periodic boundary conditions, cell lists are more common because they adapt to the existence of clearly defined coordinate extrema, necessary for the partitioning of the space into cells.

The implementation of a cell list algorithm is relatively straightforward and provides a massive speedup relative to computing all pairwise distances. However, achieving cutting-edge performance in such an implementation, particularly in the presence of periodic boundary conditions and aiming the use of parallel processing, demands specialized methods [2–4]. Performant cell list algorithms are, of course, implemented in all major particle simulation packages [4–10], and also in astrophysical simulation and analysis tools [11]. These implementations are highly specialized for computing intermolecular potentials and forces, ensuring high-performance computing speed and scalability. However, being written in low-level languages and integrated into the framework of the application, are not accessible for easy customization and reuse.

*CellListMap.jl* aims to provide a customizable, yet fast and easy-to-use, implementation of cell lists for custom particle simulations and analyses. Written in Julia [12], it allows the user to quickly write efficient yet high-level functions to be computed between pairs of particles that are close to each other, to compute distance dependent properties. With *CellListMap.jl*, it is possible to write custom analysis routines or simulation codes in a few lines of interactive Julia code, with a performance comparable with state-of-the-art analysis and simulation tools. A comprehensive user manual is available, where examples are supplied of the computation of many typical molecular and astrophysical properties, as well as of a complete particle simulation code. The package is freely available under the MIT license, at http://m3g.github.io/CellListMap.jl.

## 2 APPROACH

The purpose of the current implementation of cell lists is to provide a framework for custom calculations of pairwise-dependent properties in particle systems. We assume that particle coordinates are available. For example, trajectories obtained with the most popular molecular simulation packages can be read with the Chemfiles suite [13]. Here we describe the basic elements of the package interface. Further examples and advanced options are described in the user manual. All the code blocks presented in this article are available as Pluto notebooks with detailed step-by-step documentation, at http://github.com/m3g/CellListMapArticleCodes.

## 2.1 Overview

The purpose of the package is to allow the efficient computation of properties dependent on the distances, within a cutoff, between particles in a 2D or 3D system. The package supports general periodic boundary conditions. Therefore, the typical steps for setting up a calculation using *CellListMap.jl* are, given the coordinates of the particles:

1. Define a function which, given the distance (or coordinates) *of a single pair of particles* updates the property to be computed. This is implemented by the user for each application.

2. Define the system spatial geometry and cutoff: In this step, the system *Box* is defined, carrying the information about periodic boundary conditions and cutoff.

3. Compute the cell lists.

4. Map the function defined in step 1 through the cell lists, obtaining the updated property.

Step 1 defines the property to be computed, as desired by the user. Steps 3 to 4 consist of calling three specific functions of the package, with minimal user intervention. Scalar properties, as potential energies, can be computed, and vector properties, as forces on each particle, can be updated using this interface. Advanced usage options allow these computations to reuse memory buffers, for optimal efficiency. In the simpler cases, steps 3 and 4 will run in parallel without further user intervention. Custom parallel splitting and reduction functions can be provided by the user if required by the properties to be computed.

## 2.2. Minimal example

We split the calculation into three steps: 1) the definition of the system geometry and cutoff; 2) the construction of the cell lists; 3) the mapping of the function to be computed into the cell lists.

A minimal illustrative working example is shown in Code 1, where the sum of the distances of random 3D particles generated in a cube of unitary sides is performed.

```
1    using CellListMap
2    x = rand(3,10^5)
3    box = Box([1,1,1],0.05)
4    cl = CellList(x,box)
5    map_pairwise( (x,y,i,j,d2,output) -> output += sqrt(d2), 0., box, cl )
```

**Code 1.** Minimal working example for the computation of the sum of distances of 100k particles in a cubic periodic box of side 1 and a cutoff of 0.05. More details in [supplementary notebook 1](supplementary notebook 1).

In the first line of Code 1 the package is loaded, and in line 2 a random set of 100k particle positions is generated. The system geometry and cutoff are set in line 3. Here we illustrate the use of a periodic cubic box of side 1.0 and a cutoff of 0.05. The cell lists are constructed in line 4, and in line 5 the function to be mapped is evaluated.

## 2. 3. The mapped function

The interface of the *map_pairwise* function assumes the definition of the function to be mapped, the initial value of the *output* variable (zero in the example) and the box and cell list. The function to be mapped has to be the structure shown in Code 2.

```
1    function f(x,y,i,j,d2,output,args...)
2        # evaluate property for pair i,j and update output variable
3        return output
4    end
```

**Code 2.** General format of the function to be evaluated for each pair of particles closer than the cutoff distance, to be passed to the *map_pairwise* function. More details in supplementary notebook 2.

Internally, *map_pairwise* calls a function with the $x$, $y$, $i$, $j$, $d2$, and *output* arguments, corresponding to the positions of particles $i$ and $j$ (following the minimum-image convention) of the input set, the squared distance between the particles, and the *output* variable, which will be updated and returned from the function. An external function defined by the user must both receive an output variable as an argument and return it, but it may or may not require the use of the positions, indexes or squared distance between the particles, depending on the property to be computed. If one or more of these parameters are not required, they can be safely ignored within the method or simply removed by using an anonymous function, as shown in Code 1. Additional inputs, such as particle masses or Leonard-Jones parameters, can be specified and delivered to the inner function via a closure.

The variable defining the output may be mutable and immutable. To conform with the Julia convention of functions ending with ! being mutating functions, *map_pairwise!* and *map_pairwise* (with the bang, or not) are aliases of the same function, which always return the resulting output value. If the output variable is immutable, its value won't be mutated, and the assignment of the result to the output needs to be explicit. In these cases, it is customary to use the *map_pairwise* (without !),

```
output = map_pairwise(f::Function, output0, box, cl)
```

where `output0` represents the initial value of the immutable output. When, on the contrary, the output is a mutable variable (an array, for example), the *map_pairwise!* version is preferred for code clarity, and the reassignment is not needed (nor recommendable), as in

```
map_pairwise!(f::Function, output0, box, cl)
```

The mapped function will be evaluated only for the pairs of particles which are within the desired cutoff. The squared distance between the particles is provided because it is precomputed and usually required for the evaluation of distance-dependent pairwise properties.

## 2. 4. Coordinates

Coordinates can be given in two or three dimensions, as matrices with dimensions *(N,M)* where *N* is the dimension of the space and *M* the number of particles, or as vectors of vectors. Usually vectors of *StaticArrays* are used to represent particle coordinates. The memory layout of these arrays is the same as that of the matrices, and they can be converted into each other by a reinterpretation of the data. The coordinates can be also provided as an array of mutable vectors, which is not optimal for performance of particle computations in general, but won't have a noticeable impact in the performance of *CellListMap* because the coordinates are copied into the cell lists with static memory layouts.

## 2. 5. The system boundary conditions, and cutoff

The construction of cell lists requires that the particles are contained within a limiting box, which most commonly is associated with the periodic boundary conditions used. *CellListMap* accepts general (triclinic) periodic boundary conditions. The properties of the system periodic box, including the cutoff for the interactions used, are defined with the *Box* constructor, as illustrated in Code 3 for cubic and triclinic boxes.

```
1    julia> box = Box([10,20,15],1.2)
2    Box{OrthorhombicCell, 3, Float64, Float64, 9}
3       unit cell matrix = [ 10.0, 0.0, 0.0; 0.0, 20.0, 0.0; 0.0, 0.0, 15.0 ]
4       cutoff = 1.2
5       number of computing cells on each dimension = [10, 18, 14]
6       computing cell sizes = [1.25, 1.25, 1.25] (lcell: 1)
7       Total number of cells = 2520
8
9    julia> box = Box([ 10   0   0
10                        0  10   0
11                        0  20  10 ], 1.2)
12   Box{TriclinicCell, 3, Float64, Float64, 9}
13      unit cell matrix = [ 10.0, 0.0, 0.0; 0.0, 10.0, 20.0; 0.0, 0.0, 10.0 ]
14      cutoff = 1.2
15      number of computing cells on each dimension = [11, 11, 27]
16      computing cell sizes = [1.2, 1.2, 1.2] (lcell: 1)
17      Total number of cells = 3267
```

**Code 3.** Initialization of the system Box, with orthorhombic or triclinic periodic boundary conditions. The system's geometry is defined by the type of unit cell matrix, and an orthorhombic cell is assumed if a vector of box sides is supplied. More details in [supplementary notebook 3](#).

## 2. 6. Construction of the cell lists

The cell lists are constructed with the *CellList* constructor, illustrated in Code 4. The output of which will display the number of particles of the system, the number of cells with real particles, and the total number of particles in the computing box.

```
1    julia> cl = CellList(x,box)
2    CellList{3, Float64}
3      100000 real particles.
4      512 cells with real particles.
5      190865 particles in computing box, including images.
```

**Code 4.** Computing the cell lists from the coordinates, *x*, and the system *box*. Particles are replicated at the boundaries to avoid coordinate wrapping in the function mapping step. More details in supplementary notebook 4.

On the construction of the cell lists, the *real* particles are replicated to create ghost cells around the boundary conditions which respect the periodicity and guarantee that real particles will interact with the correct number of neighbors. This strategy allows the computations of the function mapping to be completely agnostic to the periodic boundaries, and no coordinate wrapping is needed after this step. It turns out that this particle replication is compensatory for performance [14,15]. The number of ghost particles is dependent on the geometry of the box and on the cutoff. For very large systems, it is generally expected that the cutoff is much smaller than the system size, and thus this additional memory requirement should not be the limiting factor for the computations.

A typical particle set used in the computing step is shown in Figure 1. The *real* particles are shown in green, and the ghost particles in light red.

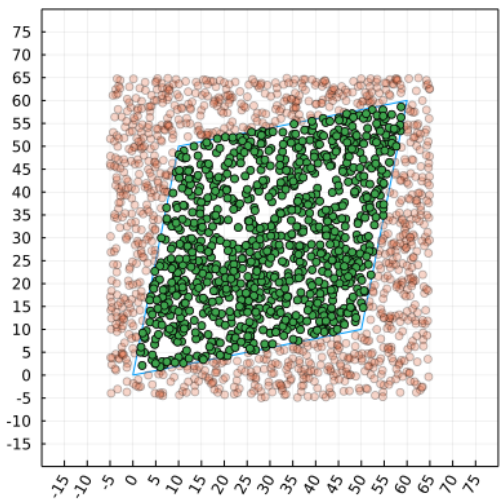

**Figure 1.** Typical computing box with a triclinic cell, in two dimensions. The *real* particles are depicted in green. Ghost particles (light red) are generated such that coordinate wrapping and minimal image calculations are not necessary in the function mapping step.

The cell lists carry a copy of the coordinates, which are stored in independent vectors for each cell. This increases the locality of the memory accesses relative to alternative (linked-list) approaches, accelerating the computation.

### 2. 7. Mapping the pairwise computation

Given the box and the cell list, any custom function can be mapped into the pairs that satisfy the distance cutoff. The user needs to define the function to be computed for one pair of particles, to update the output property. This function is then passed as a parameter to the *map_pairwise* or

*map_pairwise!* functions. The mapped function may require additional data, in which case it will *close over* the data, as shown in Codes 2 and 5. Additional examples illustrating the full flexibility of the implementation, and resulting performance are described in Section 3 and in the user manual.

```julia
1   julia> u = map_pairwise(
2              (x,y,i,j,d2,u) -> u += 1 / sqrt(d2),
3              0., box, cl
4          )
5
6   julia> const mass = rand(N)
7          u = map_pairwise(
8              (x,y,i,j,d2,u) -> u += mass[i]*mass[j] / sqrt(d2),
9              0., box, cl
10         )
```

**Code 5.** Mapping the computation of a property into the pairs of particles which, according to the cell lists and system box properties, are within the desired cutoff. In the first example the sum of the inverse of the distances is computed. In the second example, masses are used to compute a putative central potential. More details in supplementary notebook 5.

At line 2 of Code 5 the function to be mapped is defined. The interface requires 5 arguments: the coordinates of the two particles, their indices, their squared distance, and the output to be computed. Here, $u$ is a scalar, which is initialized at zero at line 3. The function in line 2 must be read as: given the input parameters of the left, return the result of the computation on the right, and is the standard Julia anonymous function syntax. Here, the actual computation only requires the squared distance between the particles, and the remaining arguments are ignored. In the example in lines 6 to 10 the masses are closed over by the anonymous function and are used for the computation of a potential energy. The closed-over variables must not be modified, because concurrent access can occur when using multi-threading.

At the same time, the implementation of this mapped function should be, for the user, agnostic to the use of multi-threaded code. Internally, the output variable to be updated will be copied and updated without concurrency by each thread, to be finally reduced when the computation is finished. This does not cause memory issues if the output variable to be updated is a scalar, but has to be taken into consideration if an array (for instance with length equal to the number of particles, as an array of forces) has to be updated. In memory-restricted scenarios this can be a limiting factor for the use of the package. Also, the preallocation of these large threaded buffers may be important for the performance of iterative computations. Preallocation is possible by passing an optional *output_threaded* parameter to the *map_pairwise* function.

## 2. 8. Current implementation details

The basic implementation of a cell list algorithm consists in partitioning the space into cells, more simply with a side equal to the cutoff, and by integer division of the coordinates of each particle by the cutoff, assigning to which cell each particle belongs. Then, a loop over the cells is performed, and the interactions of the particle of each cell with the particles of the neighboring cells are computed. In 2D each cell has 8 neighboring cells, and in 3D each cell has 26 neighboring cells. If the size of the system is much greater than the cutoff, the number of distances computed is drastically reduced.

In order to obtain a cutting-edge cell list implementation and function mapping, many improvements over the most simple algorithm are necessary. The following strategies were implemented up to version 0.7.13 of *CellListMap.jl*. These are implementation details, such that improvements on each of the methods is possible in future versions:

*Optimizations for cell list construction and pass-through*

Ghost particles are created (see Figure 1) at the boundary of the periodic system, to avoid wrapping coordinates in the mapping phase of the calculation. This also avoids the necessity of checking if the cell is in the boundary of the system and corresponding computation of minimum-image of coordinates in the hot loops. This removes branches in the code and increases the locality of the computations. This strategy requires some additional memory, but is known to be important for optimal performance in similar computations [15].

For each neighboring cell, the particle positions are projected on the axis connecting cell centers, following the method proposed by Willis et al. [16] or similarly by Gonnet [17]. The distances are computed only for the pairs of particles for which the projected distances satisfy an appropriate condition associated with the cutoff. Unlike these reference implementations, we opted to perform a partial sorting (partitioning) of the distances along the projected axis, instead of a complete sorting. The partition is performed on local structure data carrying the particle coordinates of the cells involved, and needs to be carried over for each particle of the reference cell, but it is compensatory because it requires a $O(n)$ pass on the array of particles of the neighboring cell.

Additionally, the side of the cells can be tuned to be any integer fraction of the cutoff (the *lcell* parameter of the *Box* constructor). This reduces the number of unnecessary distance computations at the expense of running over additional neighboring cells. For typical molecular densities having cells with a side half of the cutoff (*lcell=2*) is usually the best choice [18–20], although the projection of coordinates mentioned above reduces the impact of this parameter on the number of computations.

For Orthorhombic cells, only half of the neighboring cells are evaluated, to avoid the repeated distance computations for symmetric particle pairs, and the computations run over cells containing real particles only. Thus, the algorithm does not scale badly for inhomogeneous density systems, with possibly many empty cells.

*Optimizations on memory access*

In the cell list structures, the coordinates are stored as static arrays, which allow all the computations to be non-heap-allocating in the mapping function. If the mapped function is non-allocating, the full mapping won't be allocating either, except for auxiliary variables associated with multi-threading. We use a vector of lists to contain the particles on each cell (not the indexes, but an immutable copy of the particle coordinates and index). This avoids the use of linked lists, reducing non-sequential memory accesses in the hot computation loops.

*Considerations on the current parallelization strategy*

The parallelization is performed by spawning asynchronous tasks to which fractions of the number of the cells with real particles are attributed. The computations associated to each cell are assigned to threads in an alternating fashion, such that neighboring cells are associated to different threads. This improves the load distribution in the cases where the particle distribution is inhomogeneous. The number of tasks may be greater than the number of available cores or threads, and because the tasks are initialized on any available thread, this minimizes overheads associated with the inhomogeneous distribution of computations. The number of tasks can be tuned by the user to fit each specific problem (*nbatches* parameter of the *CellList* constructor).

To support the computation of general pairwise-dependent properties without transferring to the user the responsibility of dealing with concurrent access to the output variables, the output is copied for each task. Therefore, if the parallelization is done by splitting the computation into $N$ threads, the output will internally be mapped into a vector of $N$ copies of the output, which will be updated independently by each thread. By default, the reduction of the result consists of the sum of the output of each thread, however custom reduction functions can be provided. Copying the output for each thread can be memory-limiting if dealing with output arrays for very large systems, but is efficient relative to the use of atomic operations or locks, and is probably the best alternative for most applications.

## 3 EXAMPLES

Here we provide small code snippets illustrating the flexibility and user-friendly interface of *CellListMap.jl* for the computation of different common pairwise particle properties. More examples are available and will be continually updated in the user guide. In the following examples, we use some auxiliary functions to generate toy problems, which are also implemented in *CellListMap*, notably the *CellListMap.xatomic* and *CellListMap.xgalactic* functions, which generate particle coordinates and a cell with boundary conditions that mimic densities of typical molecular condensed phase systems or astrophysical galaxy distributions.

### 3. 1 Computing Lennard-Jones potential energy and forces

Computing intermolecular potential energies and forces is common in molecular simulation and analysis software. Here we illustrate how a Lennard-Jones potential can be computed with the *CellListMap* interface. The energy is a scalar and the forces are mutable arrays, such that these examples illustrate rather generically how these two types of output data have to be handled. An efficient computation of a Lennard-Jones energies and forces usually requires the decomposition of the exponential operations into smaller powers. Here, for simplicity of the codes, we use the *FastPow.jl* package that performs such decomposition though the *@fastpow* macro.

The computation of a Lennard-Jones potential for 3 million particles in 3 dimensions is illustrated in Code 6. The example resembles the minimal example in Code 1, except that here we implement the *ulj* function separately, and we pass the ε and σ parameters (of Neon) to the function that computes the energy by closing over the values in the anonymous function definition within the call to *map_pairwise.*

```
1    using CellListMap, FastPow
2    ulj(d2,ε,σ,u) = @fastpow u += 4ε*((σ^2/d2)^6 - (σ^2/d2)^3)
3    side = 31.034; cutoff = 1.2
4    x = side*rand(3,3_000_000)
5    box = Box([side,side,side],cutoff)
6    cl = CellList(x,box)
7    const ε, σ = 0.0442, 3.28 # Neon
8    u = map_pairwise((x,y,i,j,d2,u) -> ulj(d2,ε,σ,u), 0., box, cl)
```

**Code 6.** Calculation of a simple Lennard-Jones potential energy of 3 million Neon atoms in 3 dimensions, in a periodic cubic box with sides of 31.034 nm and a cutoff of 1.2 nm. More details in [supplementary notebook 6](#).

In Code 7 we provide an example of the computation of pairwise forces, where the output variable is a mutable array. We use in this case the *map_pairwise!* syntax, to indicate mutation. The computation of the forces require the identification of the particles and the knowledge of their relative position in space. Thus, the vector connecting the particles is computed in line 3, and the update of the forces vector occurs in lines 5 and 6. Note that since static arrays are used for the representation of the coordinates of the particles, the function is non-allocating.

```
1    using CellListMap, FastPow
2    function flj(x,y,i,j,d2,ε,σ,f)
3        r = y - x
4        @fastpow dudr = -12*ε*(σ^12/d2^7 - σ^6/d2^4)*r
5        f[i] = f[i] + dudr
6        f[j] = f[j] - dudr
7        return f
8    end
9    function computef(x,box)
10       cl = CellList(x,box)
11       f = zero(x) # vector similar to x but with zeros
12       ε, σ = 0.0442, 3.28 # Neon
13       map_pairwise!((x,y,i,j,d2,f) -> flj(x,y,i,j,d2,ε,σ,f), f, box, cl)
14       return f
15   end
16   x, box = CellListMap.xatomic(10^6)
17   computef(x,box)
```

**Code 7.** Example code for the calculation of a vector of forces between particles. The function will update the *f* vector. Line 16 is only to generate a set of one million particles with a typical molecular density. More details in [supplementary notebook 7](#).

An important remark about Code 7 is that, if Julia is started with multi-threading, this code will automatically run in parallel, and the user doesn't need to modify the *flj* function to account for the possible concurrent access of the force vectors. This is because, as explained in the methodological details, each thread will update an independent copy of the force vector, which will be summed up to update the output force array.

### 3. 2 Parallel computation of a k-nearest-neighbor list

All the examples shown up to now run in parallel without further intervention from the user, except for starting Julia with multithreading. Many options to improve the performance of parallel runs are available and described in the user manual. Here, we focus on the fact that some parallel computations require custom reduction functions. In the example of Code 8 we develop a (cutoff-delimited) k-nearest neighbor code, which can run in parallel.

```julia
using CellListMap
# Update a list of closest pairs
function replace_pair!(list,i,j,d2)
    pair = (i=i,j=j,dsq=d2)
    ipos = searchsortedfirst(list, pair, by = p -> p.dsq)
    if ipos <= length(list)
        list[ipos+1:end] = list[ipos:end-1]
        list[ipos] = pair
    end
    return list
end
# Custom reduction function
function reduce_list(list,list_threaded)
    for lst in list_threaded, pair in lst
        replace_pair!(list,pair...)
    end
    return list
end
x = rand(3,100)
y = rand(3,1000)
list = [ (i=0,j=0,dsq=+Inf) for _ in 1:5 ]
box = Box([1,1,1],0.1)
cl = CellList(x,y,box)
map_pairwise!(
    (x,y,i,j,d2,list) -> replace_pair!(list,i,j,d2),
    list, box, cl,
    reduce=reduce_list
)
```

**Code 8.** Example code for the calculation of a nearest-neighbor list between two independent sets of particles. A custom reduction function is required to merge lists, keeping the minimum distances. More details in [supplementary notebook 8](#).

Code 8 illustrates many important characteristics of the *CellListMap.jl* interface. First, we compute the nearest neighbors between two independent sets of coordinates (*x* and *y*), and thus in line 23 we introduce the syntax for the construction of a cell list from two sets of coordinates. The cell list will be built for the smaller set by default. To parallelize the construction of the neighbor list, a fraction of the cells is analyzed in each thread, and independent lists are built without concurrency. The merging of the lists implies checking which are the smaller distances between the threaded lists. Thus, a custom reduction function is necessary, and is implemented in lines 13 to 18. The custom reduction function is provided as an optional keyword parameter to the *map_pairwise!* function, in line 27 of the code.

### 3. 3 Type propagation: units, uncertainties, and differentiability

Julia allows implementation of generic functions rather simply, and variable types can be propagated through the code. This propagation of types allows, for example, for the automatic differentiation of Julia code. *CellListMap.jl* was written with those capabilities, and given that it is implemented with physics and chemistry problems in mind, we exemplify how units, uncertainties, and automatic differentiation can be used. The type system also allows the use of floating points of any precision or other custom defined types whenever the proper arithmetic is defined for them. Code 9 displays simple examples of these features. We use in this example the *Unitful* package for definition of units, the *Measurements* package for uncertainty propagation [21] and the *ForwardDiff* package for automatic differentiation [22]. Each of the three examples illustrate how the properties decorated with the specific type systems are propagated.

In lines 3 to 7 we show that by providing coordinates and box properties with units, the output of the computation of the sum of the distances is obtained also with the correct units. In lines 9 to 13 we create a set of coordinates with uncertainties, which are propagated through the computation of the sum of the distances, resulting in an uncertainty of the result (line 13). Finally, in lines 15 to 27 a function to compute the sum of the distance is automatically differentiated relative to each coordinate.

```julia
1    julia> using CellListMap, Unitful, Measurements, ForwardDiff
2
3    julia> x = rand(3,1000)u"nm";
4           box = Box([1.,1.,1.]u"nm",0.05u"nm")
5           cl = CellList(x,box)
6           map_pairwise((x,y,i,j,d2,out) -> out += sqrt(d2), 0.0u"nm", box, cl)
7    10.446132891923723 nm
8
9    julia> x = [ [rand()±0.1 for _ in 1:3 ] for _ in 1:1000 ]
10          box = Box([1±0, 1±0, 1±0],0.05±0.0);
11          cl = CellList(x,box);
12          map_pairwise((x,y,i,j,d2,out) -> out += sqrt(d2), 0. ± 0., box, cl)
13    9.6 ± 2.3
14
15   julia> function sum_d(x::Matrix{T},sides,cutoff) where T
16              box = Box(T.(sides),T(cutoff))
17              cl = CellList(x,box)
18              return map_pairwise(
19                  (x,y,i,j,d2,out) -> out += sqrt(d2),
20                  zero(T), box, cl
21              )
22          end
23          ForwardDiff.gradient(x -> sum_d(x,sides,cutoff), rand(3,1000))
24   3×1000 Matrix{Float64}:
25    -1.4232     1.42772    0.777501  ...    0.298885  -0.608185   1.70248
26     0.766411  -0.31754    1.28612          -0.233791   0.918614   0.000970465
27    -1.0084     0.416327  -0.328807          0.375106  -1.70196    0.470329
```

**Code 9.** Units (lines 3-7), uncertainties (lines 9-13) [21], and automatic differentiation (lines 15-27) [22] propagating through pairwise computations with *CellListMap.jl*. More details in [supplementary notebook 9](supplementary notebook 9).

# 4 PERFORMANCE

Here we illustrate the performance of custom calculations using *CellListMap* in comparison with popular and cutting-edge implementations of equivalent calculations performed by specialized software.

The benchmarks described in this section were run with *CellListMap.jl* version 0.7.2 and *NearestNeighbors.jl* version 0.5.0 in Julia 1.7.0; *scipy* version 1.3.3 and *halotools* version 0.7 within ipython3 version 7.13.0; and *NAMD 2.14-Multicore*. The comparison of neighbor list algorithms was performed on a personal computer with 4 Intel(R) Core(TM) i7-8550U CPU @ 1.80 GHz (8 threads available) and 16GB of RAM. The comparison with *halotools* was performed in a cluster compute node with 16 Intel(R) Xeon(R) CPU E5-2650 v2 @ 2.60 GHz CPUs, which is able to run 32 processes through multi-threading. The benchmarks comparing *CellListMap.jl* to *NAMD* were run in a compute node with 2 AMD EPYC 7662 processors (128 cores) and 512GB of RAM.

## 4. 1. Computation of neighbor lists

Several implementations of methods for obtaining neighbor lists are available. Most can be used to obtain the lists of pairs of particles within a cutoff, and then be used for the computation of pairwise properties. In the context of the use of *CellListMap.jl* this is suboptimal, because for most computations the list of pairs is not explicitly needed, and with *CellListMap.jl* the computation of the properties of interest can be computed directly. Nevertheless, since neighbor list algorithms are very general and commonly used for these applications, we implemented a *CellListMap.neighborlist* list function and compared the resulting performance with two important implementations of tree-based algorithms available in Julia and Python. We do not consider periodic boundary conditions in these neighbor list calculations, and for *CellListMap.jl* this actually carries the additional cost of computing the extrema of the distribution of points to set a bounding box large enough to avoid periodic interactions within the desired cutoff.

The comparison is performed against the *NearestNeighbors.jl* Julia package [23] and with the Python *scipy.spatial.cKDTree_query_ball_point [24]* implementation of the tree algorithms. We take advantage of this last comparison to illustrate that *CellListMap.jl* can be used from within Python with minimal overhead using *JuliaCall* (see the corresponding manual section at https://m3g.github.io/CellListMap.jl/0.7/python/) [25].

The benchmark consisted in computing all pairs within a cutoff of particles of two disjoint sets. The smaller set size varied between 10 and 100,000 particles, while the largest set had 1,000,000 particles. The distance trees and cell lists are constructed for the smaller set, and the pairs are constructed by running over the particles of the largest set. This provides the best performance for the range of set sizes studied here. The density of the systems was always 100 particles/nm$^3$, the atomic density in water, and the cutoff was set to 1.2 nm. Thus, the number of distance computations is similar to that of a typical molecular condensed-phase system.

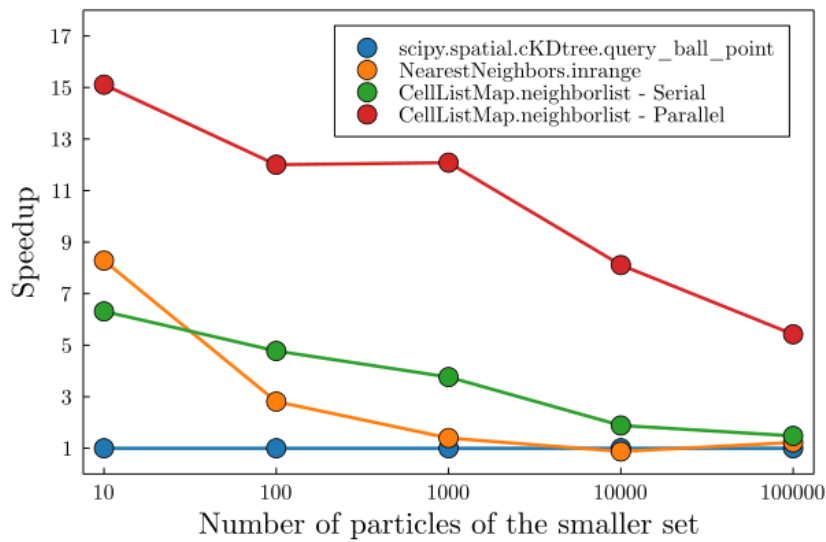

**Figure 2.** Construction of in-range neighbor lists. A cutoff of 1.2 nm for a system with water-like atomic density was used in all cases. The pairs of particles closer than the cutoff were probed for two sets, the greatest one with 1 million particles, and the smaller one with a variable number of particles from 10 to 100.000. The construction of distance trees is more expensive than that of cell lists, and for the typical density of molecular systems *CellListMap.jl* is frequently faster than the tree-based methods. Parallel runs of *CellListMap.neighborlist* were executed in a 4-core (8 threads) Intel(R) Core(TM) i7-8550U CPU @ 1.80 GHz notebook, with Julia being started with the *-t8* command-line option.

Figure 2 compares the performance of each implementation of neighbor lists. The baseline is python *scipy.spatial.cKDtree.query_ball_point* performance. The performance of the other implementations is shown as the speedup relative to this one. Both the serial and parallel implementations in *CellListMap.jl* can provide significant speedups relative to the alternatives in this setup, depending on the system size. Therefore, the package can be an interesting alternative to neighbor list computation. An important note is that the relative cost of computing the cell lists vs. distance trees can vary widely depending on the distribution and number of points. Cell lists are faster for roughly homogeneous distributions and cutoffs much smaller than the system size.

**4.2 Atomistic simulations: comparison with NAMD**

NAMD is one of the central packages of the molecular dynamics simulation ecosystem [9], and regarded as having high performance and scalability. Here, we compare the performance of a simulation of a Neon fluid where interactions are computed using *CellListMap* with a similar simulation performed with *NAMD*. Several remarks are required for the appreciation of this comparison, which aims only to compare the implementation of the computation of short-ranged interactions in the two packages: 1) The particles in the simulation only interact through Lennard-Jones potentials, thus no charges are involved. 2) The computation of long-ranged electrostatic interactions is turned off in NAMD using the "PME off" keyword option. 3) We force NAMD to update the verlet lists at every integration step, which is not the default choice for an optimally performant simulation in practice. 4) NAMD can be run on GPUs, while *CellListMap.jl* still lacks a GPU port. Still, with those considerations, the present comparison does not consider the possible additional overhead in NAMD associated with the many other features it has implemented and, on the other side, *CellListMap.jl* is a general-purpose, customizable implementation of cell lists, and not a specialized MD simulation software as *NAMD*.

The systems simulated consist of Neon fluids, simulated with CHARMM parameters [26], with a density of 100 atoms per nm$^3$, corresponding to the atomic density of liquid water. Simulations with 10k and 100k particles were performed with orthorhombic periodic boundary conditions, a cutoff of 1.2 nm for Lennard-Jones interactions (no switching was used), and with temperature control through velocity rescaling at every 10 steps. A time-step of 1 fs was used in the Velocity-Verlet method to propagate the trajectory. The complete simulation code is a ~150 lines of code of Julia, available at https://github.com/m3g/2021_FortranCon/tree/main/celllistmap_vs_namd. The code includes the definition of parameters, the implementation of the Lennard-Jones force and energy functions. Nothing was parallelized in this code except the pairwise calculations which are delegated to *CellListMap*. The simulations of 10k Neon particles were run for 5000 steps, and the simulations with 100k particles were run for 1000 steps.

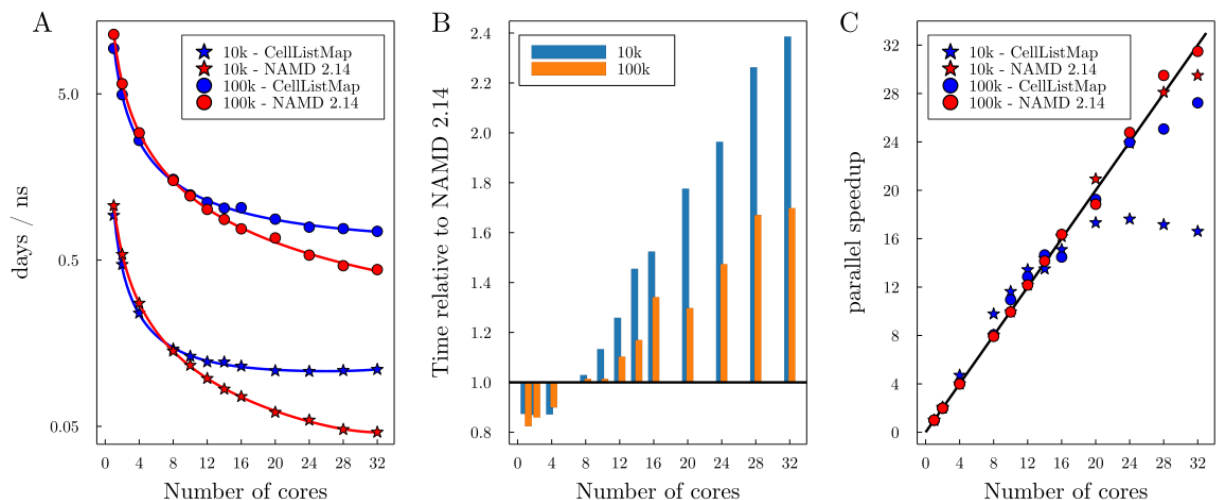

**Figure 3.** Performance of a simulation of a Neon fluid performed with NAMD 2.14, compared with a similar simulation in which non-bonded interactions were computed with *CellListMap*, both with pairlist update at every step. Simulations with 10k and 100k particles were performed. In (A) and (B) we see that the simulations performed with both implementations have similar performances for 1 to about 10 processors, and that NAMD becomes faster for a larger number of cores. Good scaling is observed for up to 20 threads for both system sizes.

Figure 3 shows that the performance of the simulation using *CellListMap* is comparable to that of NAMD. The codes run at similar speeds with about 10 threads, and NAMD displays better scaling. In these examples, however, NAMD was never much more than a factor of 2 faster than the simulation performed with *CellListMap.jl*. Thus, the implementation of cell lists available in this package is performant enough for the customized computation of short-ranged forces and other properties, in particular for the development of custom simulation analysis tools or flexible and customizable simulation codes, since in actual simulations the updating of the cell lists is not performed on every step. This strategy is indeed implemented in *Molly.jl* [27], which is a central package in the Julia molecular simulation ecosystem.

The bottlenecks of the scaling of the *CellListMap* computation, relative to *NAMD*, are of different natures: Most importantly, the construction of the cell lists does not scale very well (see Section 4.4), and becomes a limiting factor when the time required for this step becomes comparable with the time required for computing the forces. This inevitably occurs when many threads are available for computation, and particularly for smaller systems, where the computation of the forces is relatively cheap. The algorithm for the threaded construction of the lists by *NAMD* is certainly better, and this is a possible future improvement to the package. Also, we did not parallelize any other computation in this simulation, meaning that the update of positions and velocities, and velocity rescaling, are performed in serial. When the cost of these operations becomes comparable to the cost of the force update (smaller systems and/or many threads available), the scaling is penalized by a constant factor.

## 4. 3 Computing astrophysical galaxy pairwise velocities

A typical calculation in the field of astrophysics is that of relative velocities of galaxies as a function of their distances. Some packages, like *halotools* [11], implement a function to compute this distribution given the vectors of particle positions and velocities. The implementation of this computation in *halotools* is in *Cython* and not easily customizable.

Code 10 shows the complete implementation of the computation of a distance-dependent pairwise velocity distribution with *CellListMap.jl*. We only need to define a function that updates the histogram given the distance and relative velocity between a pair of particles. Since we aim to obtain the distribution of velocities as function of the distances between the galaxies, the velocities are closed over in the anonymous function definition, in line 18. The *binstep* is also closed over in that definition.

```julia
1    using CellListMap, StaticArrays
2    using LinearAlgebra: norm
3    function up_histogram!(i,j,d2,vel,binstep,hist)
4        bin = Int(div(sqrt(d2),binstep,RoundUp))
5        if bin <= size(hist,1)
6            hist[bin,1] += 1
7            hist[bin,2] += norm(vel[i] - vel[j])
8        end
9        return hist
10   end
11   function pairwise_velocities(N)
12       hist = zeros(10,2)
13       pos, box = CellListMap.xgalactic(N)
14       vel = [ rand(SVector{3,Float64}) for _ in pos ]
15       binstep = 0.5 # cutoff is 5.0
16       cl = CellList(pos,box)
17       map_pairwise!(
18           (x,y,i,j,d2,hist) -> up_histogram!(i,j,d2,vel,binstep,hist),
19           hist, box, cl,
20       )
21       return hist[:,2] ./ hist[:,1] # return average per bin
22   end
23   pairwise_velocities(10^6) # run
```

**Code 10.** Computing a histogram of average pairwise velocities between galaxies, as a function of their relative distances, a typical calculation in astrophysical simulations. More details in supplementary notebook 10.

The *CellListMap.xgalactic* is an auxiliary test function that generates a set of coordinates and a *Box* with a cutoff with dimensions typical of those of astrophysical calculations. The number of particles can be defined as an input parameter of the *pairwise_velocities* function. The histogram being updated contains on the first column the number of pairs found with distances in each bin, and in the second column the sum of the pairwise velocities. The average velocity of the pairs is computed in the last line of the *pairwise_velocities* function, to be returned (line 21).

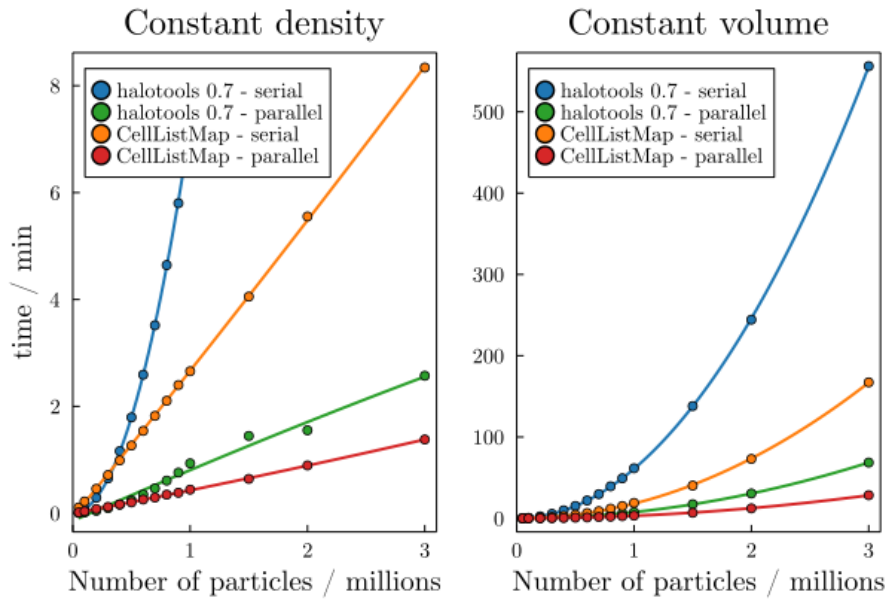

**Figure 4.** Performance for the calculation of the pairwise velocities between "galaxies", a typical calculation in the field of astrophysics, compared with the *halotools* package. In the "Constant density" panel, the density of the system and the cutoff correspond to the experimental universe galaxy density for a cutoff of 5 megaparsecs. If the density is increased, as shown in the "Constant volume" panel, the scaling is quadratic because the number of interparticle distances effectively increases. *CellListMap.jl* compares favorably with *halotools* v0.7 in both settings. These benchmarks were run in a compute node with 16 Intel(R) Xeon(R) CPU E5-2650 v2 @ 2.60 GHz CPUs; 32 threads were used for all runs.

Figure 4 compares the performance of the code shown in Code 10 relative to the *halotools* implementation of the same histogram computation. The computation based on *CellListMap.jl* performs favorably, even though it is not a specialized code for this specific calculation. Also, we were not able to run larger problems with the *halotools* implementation because of apparent memory limitations. With *CellListMap*, we were able to execute the code in this example with 100 million particles, and required ~70% of the available computer memory, for a running time of ~30 minutes.

## 4. 4. Scaling

Here we illustrate the dependence of the computational time required for a pairwise calculation as a function of the number of particles and with the number of processors used, in a shared memory architecture (currently *CellListMap* doesn't support other types of parallelism). The computation under study is shown in Code 6, and consists of the calculation of a simple Lennard-Jones potential, typically found in molecular simulations. The density of the system in the examples is constant and equal to 100 particles per $nm^3$, which is the atomic density of water.

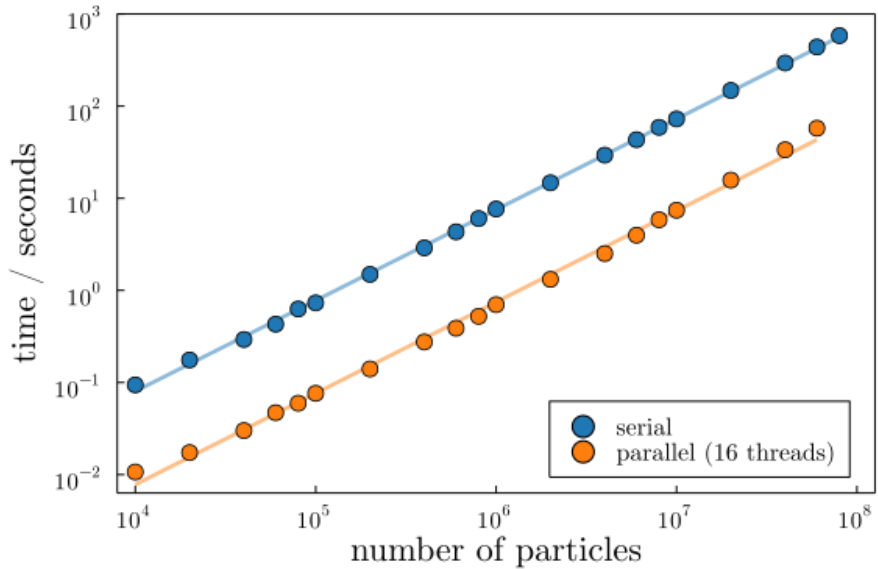

**Figure 5.** Computing time as a function of the number of particles, for systems with constant density. The scaling is linear for serial or parallel runs. These benchmarks were run in a computing node with 32Gb of RAM, which allowed the execution of the code with a maximum of 80 million and 60 million particles for the serial and parallel versions, respectively.

Figure 5 shows the time dependence of the serial and parallel versions of the computation of the Lennard-Jones potential as a function of the number of particles. From this perspective, the scaling of the package is good, being strictly linear. Parallelization requires some auxiliary memory, such that the maximum size of the problem accessible is somewhat smaller.

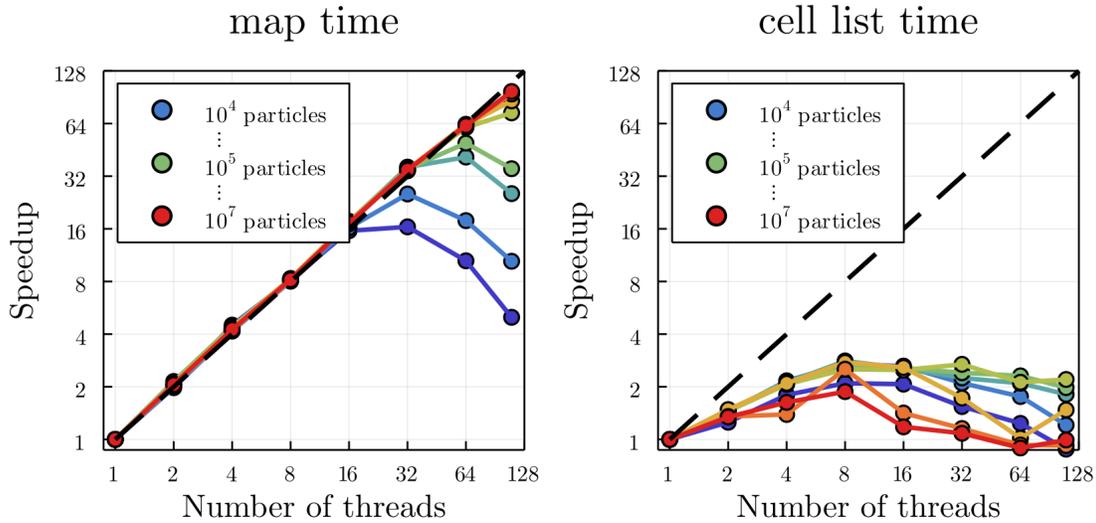

**Figure 6.** Scaling of the computation of a Lennard-Jones potential as a function of the number of cores. The dashed black line corresponds to linear scaling.

In Figure 6, on the other hand, we show the scaling of the Lennard-Jones calculation as a function of the number of threads. We split the execution into two phases: the time required for mapping the function, and the time required for the construction of the cell lists. Clearly, the mapping scales better than the construction of the cell lists, and linear scaling with up to 128 threads can be obtained for large enough systems. On the other hand, the scaling of the construction of the cell lists is not good, and achieves a maximum performance at about 8 threads, mostly independently of the number of particles. Because of this, by default the number of threads used for the construction of the cell lists is at most 8, and the number of threads used for the mapping phase is limited to 32 for smaller systems. These parameters can be tuned by the user.

Concerning the example in Code 6, the time required for the construction of the cell lists, without multithreading, is a tenth of the time required for mapping the Lennard-Jones potential on the pairs. Thus, the good scaling of the mapping phase is reflected into the overall performance of the calculation for a smaller number of threads. When the number of threads is greater, the bottleneck can be the construction of the cell lists. Further improvements, particularly on the cell list construction phase, are necessary. The relative importance of the cell list construction is dependent, of course, on the cost of the function being mapped and the total number of cores available. Typically the mapping phase is more expensive than the construction of the cell lists. If the function being mapped on pairs is expensive, if the cutoff is larger, or if the number of parallel threads is limited, it is typical that the bad scaling of the cell list construction is not relevant for the total computation time.

## 5 CONCLUSION

Here we present an implementation of cell lists in Julia, to be used in the development of custom simulation and trajectory analysis programs. The implementation is designed in such a way that it is simple to write small programs that can quite efficiently compute pairwise dependent properties, for the particles of a system within a cutoff. The code is performant, comparable to cutting-edge packages for computing neighbor lists, simulations and other $n$-body system properties in molecular and astrophysical simulations. Future developments may include the improvement of the performance of the cell list construction phase and the implementation of GPU-accelerated or distributed-computing versions. The package is freely available at http://m3g.github.com/CellListMap.jl, and is already used in simulation analyses [28] and production code [27].

## SUPPLEMENTARY INFORMATION

Detailed explanations of the code blocks are available as Pluto notebooks at: http://github.com/m3g/CellListMapArticleCodes.

## ACKNOWLEDGEMENTS

The author acknowledges the financial support of Fapesp (2010/16947-9, 2018/24293-0, 2013/08293-7, 2018/14274-9) and CNPq (302332/2016-2). Research developed with the help of CENAPAD-SP (National Center for High Performance Processing in São Paulo), project UNICAMP / FINEP - MCTI. The author is deeply indebted to many participants of the Julia Discourse forum, which contributed to many of the ideas incorporated into the package.

## CONFLICT OF INTEREST

The author declares that there are no conflicts of interests.